\documentclass[aps,prd,twocolumn,groupedaddress,nofootinbib,10pt]{revtex4-1}
\usepackage{tensor, amsmath, amssymb, graphics, geometry, mathtools, graphicx, wrapfig}
\usepackage[
	colorlinks=true,
	linkcolor=blue,
	citecolor=blue,
	filecolor=magenta,
	urlcolor=blue
]{hyperref}
\newcommand{\barT}{{\overline{T}}}

\begin{document}

\title{Comparison of \textit{f}(\textit{R},\textit{T}) Gravity with Multiple Datasets}
\author{Vincent R. Siggia}
\author{Eric D. Carlson}
\email{ecarlson@wfu.edu}
\affiliation{Department of Physics, Wake Forest University, 1834 Wake Forest Road, Winston-Salem, North Carolina 27109, USA}

\begin{abstract}
    We examine $f(R,T)=R+\lambda T^\epsilon$ gravity models by finding the best fit parameters for various values of $\epsilon$. This is accomplished by analyzing data from the cosmic microwave background, baryon acoustic oscillation observations, cosmic chronometer, and Type Ia supernovae introducing correlations and radiation effects to our previous work. We find the probability distribution for the best fit model around the value of $\epsilon=0.010^{+0.013}_{-0.021}$, consistent with the standard cosmological model value of $\epsilon = 0$.
\end{abstract}

\maketitle

\section{Introduction}

The Universe is accelerating as observed by Type Ia supernovae (SNe Ia) \cite{Riess_1998,Perlmutter_1998,Brout_2022,Riess_2022} and measurements of the cosmic microwave background (CMB) \cite{Planck2018,Chen_2019}. To elucidate the cause, many additional measurements, such as baryon acoustic oscillations (BAO)  \cite{DESI_DR2,10.1111/j.1365-2966.2009.15812.x,Kazin_2010,10.1111/j.1365-2966.2011.19250.x,WiggleZ,10.1093/mnras/stt988,10.1093/mnras/stu523,10.1093/mnras/stv154,10.1093/mnras/stw2373,10.1093/mnras/stx1641,refId0,10.1093/mnras/sty1955,refId1,10.1093/mnras/staa3234,10.1093/mnras/staa3050} and cosmic chronometers (CC) \cite{Zhang_2014,Jimenez_2003,PhysRevD.71.123001,M.Moresco_2012,10.1093/mnrasl/slv037,Moresco_2016,10.1093/mnras/stx301,Stern_2010,Borghi_2022,10.1093/mnras/stad1621}, have sequently been performed. The most common explanation, which is now considered the standard cosmological model, is the combination of a cosmological constant combined with cold dark matter ($\Lambda$CDM). However, many puzzles remain in this model: the Hubble tension between the CMB observations and distances from SNe Ia; and the apparent discrepancy between the observations from the CMB and the distribution of matter caused by BAO \cite{DESI_DR2}. 

Although $\Lambda$CDM has several successes, many alternatives have been considered. One alternative is $f(R)$ gravity, where the Einstein-Hilbert action is replaced by a general function of the Ricci curvature $R$ \cite{10.1093/mnras/150.1.1}. Some versions of this model have been claimed to fit the cosmological data significantly better than the $\Lambda$CDM model \cite{ODINTSOV2023137988,GameOver}. These theories can be extended further to also include terms from the stress-energy trace $T$, called $f(R,T)$ gravity \cite{Harko_2011,PhysRevD.90.028501,Fisher_2019}. In previous papers \cite{Siggia_2025,Siggia_2026}, we exclusively looked at SNe Ia data and found that nearly all values of $\epsilon$ in the range $(-\infty,0.03)$ fit the data and were competitive with $\Lambda$CDM. In this paper, we examine the same type of model $f(R,T) = R + \lambda T^\epsilon$, using updated SNe Ia data \cite{Brout_2022,Riess_2022} and also include data analyzing the CMB, BAO, and cosmic chronometers (CC), with the intention of studying the likelihood of different values of $\epsilon$. In contrast to our previous work, as we will demonstrate, including the latest supernova data, as well as CMB, BAO, and CC, strongly constrain $\epsilon$ to be quite small, and the resulting range is easily compatible with the $\Lambda$CDM value of $\epsilon = 0$.

In Sec.~\ref{sec:General}, the general framework for our models including radiation is worked out. Next, in Sec.~\ref{sec:Hubble}, the Hubble parameter is derived for our models. Then in Sec.~\ref{sec:Various}, various datasets are introduced. In particular, the details for the analysis of CMB, BAO, CC, and SNe Ia are listed. Afterwards in Sec.~\ref{sec:Stat}, the behavior and statistical analysis of the models is discussed. In Sec.~\ref{Sec:Conclusion} our results are summarized.

We will use units where $\hbar = c = 1$, and we use the mostly minus sign convention for the metric. The Riemann tensor is given by ${R^\alpha}_{\mu\beta\nu} = \nabla_\beta \Gamma^\alpha_{\mu\nu} - \cdots$ and the Ricci Tensor by $R_{\mu\nu} = {R^\alpha}_{\mu\alpha\nu}$.

\section{General Model}\label{sec:General}

In our previous papers \cite{Siggia_2025,Siggia_2026}, we ignored the contribution from radiation because we were exclusively discussing SNe Ia, which focuses on late times when the contribution from radiation is minimal. In order to analyze the CMB, BAO, CC, and SNe Ia altogether, we now want to incorporate radiation effects. Thus, our action is
\begin{equation}\label{Action}
    S = \int d^4x \sqrt{-g}\left[\mathcal{L}_m+\mathcal{L}_r -\frac{1}{2\kappa^2}\big(R+\lambda T^\epsilon\big) \right] \; ,
\end{equation}
where $\mathcal{L}_m$ is the Lagrangian for ordinary and dark matter, $\mathcal{L}_r$ is the contribution from radiation, and $\kappa^2=8\pi G$.

Varying the action Eq.~(\ref{Action}) with respect to $g^{\mu\nu}$, gives us the $f(R,T)$ Einstein equations for our model as
\begin{equation}\label{Einstein EQ}
   R_{\mu \nu }-\frac{1}{2}R\,g_{\mu \nu }=\kappa ^2T_{\mu\nu}+\frac{1}{2}\lambda T^\epsilon g_{\mu \nu }-\epsilon\lambda T^{\epsilon-1
   }\frac{\partial\,T}{\partial g^{\mu \nu }}\;,
\end{equation}
where $T_{\mu\nu}=T^r_{\mu \nu }+T^m_{\mu \nu }$ and $T^r=0$. The divergence of Eq.~(\ref{Einstein EQ}), as shown in \cite{Siggia_2025,Siggia_2026} is
\begin{equation}
   \kappa^2\nabla^\mu T_{\mu\nu}= \nabla^\mu\left(\epsilon\lambda T^{\epsilon-1}\frac{\partial\,T}{\partial g^{\mu \nu }}\right)-\frac{1}{2}\epsilon\lambda T^{\epsilon-1}\nabla_\nu T\;.
\end{equation}
Note that stress-energy is not conserved \cite{PhysRevD.90.028501,Harko_2011}. As discussed in \cite{Fisher_2019,Siggia_2025,Siggia_2026}, the matter Lagrangian must be written as a function of the number density $n$ and entropy per particle $s$, as well as various terms involving Lagrange multipliers in an attempt to conserve particle number and entropy. As a result, the trivial current density $J^\mu=n u^\mu$ is not conserved; rather, it is the effective current density
\begin{equation}\label{General Current Density}
   J'^\mu=\left(1+\frac{2\epsilon\lambda }{\kappa ^2}T^{\epsilon-1}\right)J^\mu\;
\end{equation}
that is conserved.

Since the stress-energy tensor is conserved in standard gravity, it can be shown that the energy density and pressure are related by
\begin{equation}
    n\frac{\partial\rho}{\partial n}=\rho+p.
\end{equation}
In general, this does not hold in $f(R,T)$ gravity. Nevertheless, this can still be treated as a definition of the naïve pressure $p$.

The stress-energy quantities appearing in Eq.~(\ref{Einstein EQ}) become
\begin{subequations}
    \begin{align}
         \label{General Stress-Energy}T_{\mu \nu}&=(\rho +p)u_{\mu }u_{\nu }-g_{\mu \nu }(\mathcal{L}_m+\mathcal{L}_r) \; , \\
        \label{General Stress-Energy Trace}T&=\rho_m +p_m-4\mathcal{L}_m\;, \\
        \label{T Variation}\frac{\partial T}{\partial g^{\mu\nu }}&=-\frac{1}{2}u_{\mu }u_{\nu }\left(4+n\frac{\partial }{\partial n}\right)(\rho_m +p_m)\;.
    \end{align}
\end{subequations}
Unfortunately, the matter Lagrangian still contains Lagrange multipliers. All quantities can be calculated ``on-shell''; chosen to satisfy the equations of motion in order to eliminate the Lagrange multiplier terms. We denote such quantities by an over-bar. Following the analysis of \cite{Siggia_2025}, eliminating these Lagrange multipliers leads to the on-shell matter Lagrangian
\begin{equation}\label{lm-bar}
    \overline{\mathcal{L}_m}=p_m+\frac{\epsilon\lambda}{2\kappa ^2\barT^{1-\epsilon}+4\epsilon\lambda}n\frac{\partial }{\partial n}(\rho_m +p_m)\;.
\end{equation}
Then $\overline{\mathcal{L}_m}$ can be substituted into  Eqs.~(\ref{General Stress-Energy})-(\ref{T Variation}) to produce their on-shell equivalents. Similarly, by combining Eqs.~(\ref{General Stress-Energy Trace}) and (\ref{lm-bar}), the on-shell stress-energy trace $\barT$ becomes an implicit function of $\rho$ and $p$ given by
\begin{equation}\label{General Quad}
    \rho_m-3p_m=\barT+\frac{2\lambda\epsilon }{\kappa ^2}\barT^{\epsilon }+\frac{2\lambda\epsilon }{\kappa ^2}\barT^{\epsilon -1}\left(4+n\frac{\partial }{\partial n}\right)p_m\;.
\end{equation}
In contrast to the matter contributions, the radiation is much simpler to analyze because it does not contribute to the trace $T$, and therefore satisfies the usual relation $\rho_r = 3p_r$. 

Assuming a flat Friedmann-Lemaître-Robinson-Walker (FLRW) metric, 
\begin{equation}\label{FLRW Metric}
    ds^2=d t^2-a^2(t)\left(d r^2+r^2d\Omega^2\right)\;,
\end{equation}
where the scale factor today is defined to be $a(t_0)=1$, we find the equations of motion in terms of $n$ and $s$ are
\begin{subequations}
\begin{align}
    \nonumber3\left(\frac{\dot{a}}{a}\right)^2&=\kappa^2\left(\rho_r+\frac{3\rho_m+3p_m+\barT}{4}\right)+\frac{\lambda}{2}\barT^\epsilon\\
   &+\frac{\epsilon\lambda}{2}\barT^{\,-(1-\epsilon)}\left(4+n\frac{\partial}{\partial n}\right)(\rho_m+p_m)\;,\label{tt-Friedmann}\\
    \nonumber-6\left(\frac{\ddot{a}}{a}\right)&=\kappa^2\left(2\rho_r+\frac{3\rho_m+3p_m-\barT}{2}\right)-\lambda\barT^\epsilon\\
    &+\frac{\epsilon\lambda}{2}\barT^{\,-(1-\epsilon)}\left(4+n\frac{\partial}{\partial n}\right)(\rho_m+p_m)\,.\label{rr-Friedmann}
\end{align}
\end{subequations}

\section{The Hubble Parameter}\label{sec:Hubble}

During the eras we are considering, the matter is non-relativistic. Because it is derived exclusively from the matter Lagrangian $\mathcal{L}_m$, we expect its pressure to be negligible, $p_m=0$, and the naïve matter energy-density will, as usual, satisfy $\rho_m\propto n$ \cite{Siggia_2025}. This reduces Eq.~(\ref{General Quad}) to
\begin{equation}\label{rho m}
    \rho_m=\barT+\frac{2\epsilon\lambda}{\kappa ^2}\barT^\epsilon\;,
\end{equation}
which can be solved for $\barT$. This allows Eqs.~(\ref{tt-Friedmann}) and (\ref{rr-Friedmann}) to be rewritten in terms of $\barT$ as
\begin{subequations}
\begin{align}  
    \label{1st P=0}3\,&\frac{\dot{a}^2}{a^2}=\kappa^2\rho_r+\kappa^2\barT\left[1+\frac{(8\epsilon+1)\lambda}{2\kappa^2\barT^{1-\epsilon}}+5\bigg(\frac{\epsilon\lambda}{\kappa^2\barT^{1-\epsilon}}\bigg)^2\right],\\
    \label{2nd P=0}-6\,&\frac{\ddot{a}}{a}=2\kappa^2\rho_r+\kappa^2\barT\left[1+\frac{(11\epsilon-2)\lambda}{2\kappa^2\barT^{1-\epsilon}}+5\bigg(\frac{\epsilon\lambda}{\kappa^2\barT^{1-\epsilon}}\bigg)^2\right].
\end{align}
\end{subequations}

The effective current density can be reduced using Eq.~(\ref{rho m}) to
\begin{equation}
    J'^{\mu }=\frac{\rho_m}{\barT} \left(n\,u^{\mu }\right)\;.
\end{equation}
During the period we are considering, the matter is nonrelativistic. The matter Lagrangian $\mathcal{L}_m$ is used to determine both $n$ and $\rho_m$, and we therefore expect $\rho_m\propto n$, so that $J'^0 \propto \rho_m^2/\barT$ will be conserved, which we can write as
\begin{equation}\label{Constant Conservation}
   \frac{\rho_m^2}{\barT}a^3=\frac{12N^3}{\kappa^2}\;,
\end{equation}
where $N$ is an arbitrary normalization factor. We can then use Eqs.~(\ref{rho m}) and (\ref{Constant Conservation}) to derive two equations for the scale factor $a$:
\begin{align}
    \label{A}a&=N\Bigg[\frac{12\kappa^2\barT^{1-2\epsilon}}{\big(\kappa^2\barT^{1-\epsilon}+2\epsilon\lambda\big)^2}\Bigg]^{1/3}\; , \\
    \label{DotA/A}\frac{\dot{a}}{a}&=-\frac{\dot{\barT}\big[\kappa^2\barT^{1-\epsilon}-(1-2\epsilon)2\epsilon\lambda\big]}{3\barT\big(\kappa^2\barT^{1-\epsilon}+2\epsilon\lambda\big)}\;.
\end{align}

It is useful to define dimensionless quantities $\tau$ and $x$ as
\begin{align}
    &\tau =t \sqrt{\frac{\kappa^2}{6} \left(\frac{2\lambda }{\kappa ^2}\right)^{\frac{1}{1-\epsilon}}}\; , \label{t scaling} \\
    &\barT(t)=x(\tau)\left(\frac{2\lambda }{\kappa ^2}\right)^{\frac{1}{1-\epsilon}}\;. \label{T Scaling}
\end{align}
We can write the redshift $z$ in terms of $x$ and $x_0 = x(\tau_0)$ as
\begin{equation}\label{Redshift}
    1+z\equiv\frac{1}{a}=\left[\frac{x_0^{1-2\epsilon}}{(x_0^{1-\epsilon}+\epsilon)^2}\right]^{1/3}\left[\frac{(x^{1-\epsilon}+\epsilon)^2}{x^{1-2\epsilon}}\right]^{1/3}\;.
\end{equation}

 This allows Eq.~(\ref{1st P=0}) to be written as
\begin{equation}\label{H^2 1st form}
    H^2=\frac{\kappa^2}{3}\rho_{r}+\frac{\kappa^2}{12}\left(\frac{2\lambda}{\kappa^2}\right)^{1-\epsilon}\frac{1}{\overline{\lambda}}\;.
\end{equation}
where
\begin{equation}\label{lambda}
    \overline{\lambda}\equiv\frac{x^{1-2\epsilon}}{4x^{2(1-\epsilon)}+(1+8\epsilon)x^{1-\epsilon}+5\epsilon^2} \; .
\end{equation}
By evaluating Eq.~(\ref{H^2 1st form}) today at $\tau_0$, $\lambda$ can be written
\begin{equation}
    \lambda=\frac12\kappa^{2\epsilon} \left[12(1-\Omega_r)H_0^2\overline{\lambda}_0\right]^{1-\epsilon}\;,
\end{equation}
where $\overline{\lambda}_0=\overline{\lambda}(x_0)$ and $H_0^2\Omega_r = \frac{1}{3} \kappa^2 \rho^0_r$. However, since $\lambda$ has units that depend on $\epsilon$, the unitless parameter
\begin{equation}\label{xi}
    \xi\equiv\frac{\lambda}{\kappa^{2\epsilon}H_0^{2(1-\epsilon)}}=\frac12 \left[12(1-\Omega_r)\bar{\lambda}_0\right]^{1-\epsilon}
\end{equation}
is used for numerical convenience.

It is possible to rewrite Eq.~(\ref{H^2 1st form}) solely in terms of $x$, $x_0$, and $H_0$ as
\begin{equation}
    \label{H^2 2nd form}H^2=H_0^2\left[\Omega_r a^{-4}+\left(1-\Omega_r\right)\frac{\overline{\lambda}_0}{\overline{\lambda}}\right]\;,
\end{equation}
where $a$ comes from Eq.~(\ref{Redshift}), 
\begin{equation}
    H_0^2\Omega_r=\frac{\kappa^2}{3}\frac{\pi^2(k_\text{B}\mathcal{T}_0)^4}{15}\left(1+\frac{\rho_\nu}{\rho_\gamma}\right)\; ,
\end{equation}
with $\mathcal{T}_0\approx2.7255\,\text{K}$, $\rho_\nu/\rho_\gamma\approx0.6913$ \cite{Froustey_2020,Bennett_2021,Akita_2020,Binder_2025}, and
\begin{equation}
    \label{lambda ratio}\frac{\overline{\lambda}_0}{\overline{\lambda}}=\left[1-\overline{\lambda}_0\left(\frac{x_0^{1-\epsilon}+\epsilon^2}{x_0^{1-2\epsilon}}\right)\right]a^{-3}+\overline{\lambda}_0\left(\frac{x^{1-\epsilon}+\epsilon^2}{x^{1-2\epsilon}}\right)\;.
\end{equation}
We note that the first term in this equation has the characteristic $a^{-3}$ dependence one would normally associate with matter, while the remaining term has the more complicated dependence one might associate with dark energy. Hence, we interpret the first term as $\Omega_m$, explicitly
\begin{equation}\label{Omega_m} 
    \Omega_m = (1-\Omega_r) \left[1-\overline{\lambda}_0\left(\frac{x_0^{1-\epsilon}+\epsilon^2}{x_0^{1-2\epsilon}}\right)\right] \; .
\end{equation}

\section{Various Datasets}\label{sec:Various}

Among the various datasets below, we consider five free parameters: $\epsilon$, $\xi$, $\omega_b$, $H_0$, and $M$, where $M$ is the absolute magnitude of Type Ia supernovae. The quantity $\omega_b$ is a combination of the current baryon density fraction $\Omega_b$ and the scaled Hubble parameter $h$, given by
\begin{equation}
\omega_b = \Omega_b h^2 = \Omega_b\left(\frac{ H_0}{100\,\text{km/s/Mpc}}\right)^2
\end{equation}
For each of these parameters, we calculate
\begin{equation}
    \chi^2=\chi^2_{\text{CMB}}+\chi^2_{\text{BAO}}+\chi^2_{\text{CC}}+\chi^2_{\text{SN}}+\chi^2_M\;,
\end{equation}
where the terms correspond to CMB, BAO, cosmic chronometer, SNe Ia, and $M$ respectively.

For each data source $k$, $\chi^2_k$ was calculated as
\begin{equation}
    \chi^2_k=\sum_{i,j=1}^{n_k}(C^{-1}_k)_{i j}\Delta^i_k\Delta^j_k\;,
\end{equation}
where $C_k$ is the covariance matrix and $\Delta_k$ is the difference between the data and the theoretical predictions of the models. For uncorrelated data, the covariance matrix $C_k$ is diagonal.

\subsection{The Cosmic Microwave Background}\label{sec:CMB}

Since we assume a flat FLRW metric, the line-of-sight comoving distance $D_\text{C}$ and transverse comoving distance $D_\text{M}$ are equal. Therefore,
\begin{equation}
    D_\text{M}(z)= \int_0^z\frac{d\tilde{z}}{H(\tilde{z})}=\int_{x_0}^{x}\frac{z'(\tilde{x})}{H(\tilde{x})}d \tilde{x}\;.
\end{equation}
The CMB power spectrum can be characterized by
\begin{equation}
    R=\sqrt{\Omega_m}H_0 D_\text{M}(z_*)\;,\quad\ell_\text{A}=\frac{\pi D_\text{M}(z_*)}{r_\text{s}(z_*)}\;,
\end{equation}
and $\omega_b$, where
\begin{equation}
    r_\text{s}(z)=\frac{1}{\sqrt{3}}\int_0^{1/(1+z)}\frac{da}{a^2H(a)\sqrt{1+[3\Omega_b/(4\Omega_\gamma)]a}},
\end{equation}
is the comoving sound horizon, $z_*\approx1090$ is the redshift at photon decoupling, $R$ is a “shift parameter” that influences the temperature spectrum of the CMB along the line-of-sight direction, and $\ell_\text{A}$ is an acoustic scale length that characterizes the temperature power spectrum of the CMB in the transverse direction \cite{Planck2018,Chen_2019,GameOver}. The distance priors are from \cite{Chen_2019} using the $w$CDM values, and correlations are considered.

\subsection{Baryon Acoustic Oscillations and the Dark Energy Spectroscopic Instrument}\label{sec:BAO}

Another type of data that will be used are BAO. During the early universe, the Universe was filled with photons combined with a hot, dense plasma of electrons and baryons, with scattering forcing them to flow together.  The overly dense regions tend to condense via gravitational attraction, but the photons exert outward pressure against this attraction, causing the mixture to oscillate as relativistic sound waves. These two fluids become decoupled and can start flowing separately, shortly after recombination, so that at approximately $z_\text{drag}\approx 1060$ the photons no longer drag the baryons, leaving a permanent imprint that can be measured  on the distribution of galaxies today \cite{Planck2018}. 

The model independent ratios 
\begin{equation}
    d_z(z)=\frac{r_\text{s}(z_\text{drag})}{D_\text{V}(z)}\quad \hbox{and} \quad A(z)=\frac{H_0\sqrt{\Omega_m}}{z}D_\text{V}(z)\;,
\end{equation}
where $\Omega_m$ is given by Eq.~(\ref{Omega_m}) and
\begin{equation}
    D_\text{V}(z)=\left[z \frac{D_\text{M}^2(z)}{H(z)}\right]^{1/3}\;
\end{equation}
is the spherically-averaged distance, are different measures used to study BAO. We used 9 data points from the Dark Energy Spectroscopic Instrument Data Release 2 (DESI DR2), available from \cite{DESI_DR2}, and 21 data points that were used from \cite{10.1111/j.1365-2966.2009.15812.x,Kazin_2010,10.1111/j.1365-2966.2011.19250.x,WiggleZ,10.1093/mnras/stt988,10.1093/mnras/stu523,10.1093/mnras/stv154,10.1093/mnras/stw2373,10.1093/mnras/stx1641,refId0,10.1093/mnras/sty1955,refId1,10.1093/mnras/staa3234,10.1093/mnras/staa3050}, compiled by Odintsov \textit{et al.} in \cite{ODINTSOV2023137988}. Therefore, we do not list the data here. The DESI DR2 data contain correlations for $1/d_z$ and $D_\text{M}/D_H$ where $D_H=c/H_0$. Also note that two pieces of $d_z$ data from \cite{10.1111/j.1365-2966.2009.15812.x} are correlated with $r=0.337$ and another two pairs of $A(z)$ data from \cite{WiggleZ} are correlated with $r=0.369$ and $r=0.438$. This affects the two resulting covariance matrices: one for $d_z(z)$ and one for $A(z)$. According to the suggestion of \cite{WiggleZ}, certain data were excluded to provide a more reliable cosmological fit.

\subsection{Cosmic Chronometers}\label{sec:CC}

A third source for analyzing the parameter-space is cosmic chronometers. These chronometers are essentially pairs of non-star-forming galaxies at similar redshift $z$. Due to no stars forming within these passively evolving galaxies, it is easy to determine the relative age difference $\Delta t$ between the two  by analyzing the metallicity of the two.  In turn, this can be used to approximate the Hubble parameter via
\begin{equation}
    H(z)=\frac{\dot{a}}{a}\approx-\frac{1}{1+z}\frac{\Delta z}{\Delta t}\;.
\end{equation}
The data from the 32 cosmic chronometers used are from \cite{Zhang_2014,Jimenez_2003,PhysRevD.71.123001,M.Moresco_2012,Moresco_2016,10.1093/mnras/stx301,Stern_2010,Borghi_2022,10.1093/mnrasl/slv037} and were compiled by Favale \textit{et al.} in \cite{10.1093/mnras/stad1621}.

\begin{table*}
\centering
\begin{tabular}{|c||c|c|c|c|c|c|c|c|c|} 
 \hline
 $\epsilon$ & $\xi_\text{lim}$ & $\xi$ & $\omega_b\,(10^{-2})$ & $H_0$ & $\chi^2_\text{min}$ & $\det(\alpha)$ & $r_{\xi\omega}$ & $r_{\xi H}$ & $r_{\omega H}$  \\
 \hline
 \hline
 0.028 & 4.238 & $4.240\pm0.019$ & $2.269\pm0.012$ & $68.24\pm0.21$ & 2075.03 & $8.388\times 10^{-3}$ & 0.270 & 0.933 & 0.557  \\
 \hline
 0 & 5.999 & $4.210\pm 0.016$ & $2.263\pm0.012$ & $68.95\pm0.22$ & 2076.46 & $7.751\times 10^{-3}$ & 0.262 & 0.938 & 0.541 \\
\hline
-0.030 & 5.644 & $4.244\pm 0.015$ & $2.258\pm0.012$ & $69.48\pm 0.23$ & 2079.74 & $7.518\times 10^{-3}$ & 0.258 & 0.939 & 0.536 \\ 
 \hline
\end{tabular}
\caption{Parameter minima values, uncertainties, and correlations for representative values of $\epsilon$ \label{tab:DataTableCorr}}
\end{table*}

\subsection{Type Ia Supernovae}\label{sec:SNe Ia}

SNe Ia will be analyzed as the fourth source of data. However, unlike our previous papers \cite{Siggia_2025,Siggia_2026}, the SNe Ia data will be from Pantheon+SH0ES \cite{Brout_2022,Riess_2022}. The Pantheon+SH0ES contains data from 1701 light curves of 1550 distinct Type Ia supernovae. Therefore, the correlations must be taken into account. Due to SNe Ia being at low redshift $z$, radiation effects are negligible. 

This allows to follow the analysis from \cite{Siggia_2026}, where we found 
\begin{align}\label{diff X}
    \nonumber\frac{dx}{d\tau}=-3&\frac{x^{1-\epsilon}+\epsilon}{x^{1-\epsilon}-(1-2\epsilon)\epsilon}\\
    &\times\sqrt{\frac{4x^{2(1-\epsilon)}+(8\epsilon+1)x^{1-\epsilon}+5\epsilon^2}{2x^{-(1+2\epsilon)}}}\;,
\end{align}
and
\begin{align}
    \nonumber &H_0 d_\text{L}=(1+z)\left[\frac{4x^{2(1-\epsilon)}_0+(1+8\epsilon)x^{1-\epsilon}_0+5\epsilon^2}{2x^{1-2\epsilon}_0}\right]^{1/2}\times\\
     &\;\left[\frac{x^{1-2\epsilon}_0}{(x^{1-\epsilon}_0+\epsilon)^2}\right]^{1/3}\int_\tau^{\tau_0}\left[\frac{(x(\tau')^{1-\epsilon}+\epsilon)^2}{x(\tau')^{1-2\epsilon}}\right]^{1/3}d\tau'\;.
\end{align}

The distance modulus $\mu$ can be easily calculated from the luminosity distance $d_L$, using
\begin{equation}
    \mu\equiv m-M=5\,\log_{10}\left(\frac{d_\text{L}}{10\,\text{pc}}\right)\;.
\end{equation}

In addition to analyzing $\mu$ for SNe Ia from the Pantheon+ data, we also include the absolute magnitude $M=-19.253\pm0.027$ from SH0ES compared against the minimized absolute magnitude from the entire Pantheon+ for each model \cite{Riess_2022}. This gives us the contribution $\chi^2_M$.

\section{Model Behavior and Statistical Analysis}\label{sec:Stat}

We are trying to study the calculated values of $\chi^2$ for arbitrary values of the five parameters,
\begin{equation}
    V^i=(\epsilon,\xi,\omega_b,H_0,M) \; .
\end{equation}
We are most interested in $\epsilon$, and therefore we treat the other four parameters as nuisance parameters which must be integrated over. For fixed $(\epsilon,\xi,\omega_b,H_0)$, the value of $\chi^2$ is always a quadratic function of $M$ with the same uncertainty $\sigma_M$. This allows us to trivially integrate over $M$. This, in turn, allows us to ignore this as a parameter and treat $\chi^2$ as a function of only the other four parameters.

We find for the values of $\epsilon$ considered, in the neighborhood of the most likely parameter values, $\chi^2$ is well-fit by a quadratic expression over the remaining three parameters so that
\begin{equation}\label{chi quadratic}
\chi^2 \approx \chi^2_\text{min} + \sum_{i,j=2}^{4} \alpha_{ij} (V^i-V^i_\text{min} )(V^j -V^j_\text{min}) \; ,
\end{equation}
where each of the parameters $V^i_\text{min}$ and $\alpha_{ij}$ are functions of $\epsilon$. We therefore vary $\epsilon$ and study its behavior.

For each value of $\epsilon$, we can use the approximation in Eq.~(\ref{chi quadratic}) to estimate the central value and uncertainties of the other parameters, as well as their correlations. Three representative values of $\epsilon$ and their corresponding values of $V^i_\text{min}$, their uncertainties, and their corresponding correlations appear in Tab.~\ref{tab:DataTableCorr}. 

For each $\epsilon$, we can then calculate the relative probability using
\begin{equation}
    \rho(\epsilon)\propto\int \exp(-\chi^2/2) \prod_{j=2}^4\frac{dV^j}{\sqrt{2\pi}}
\end{equation}

Note from our previous papers \cite{Siggia_2025,Siggia_2026}, if $\epsilon\in(0,1/2]$ there are values of $a$ for which there is no value of $\barT$ that satisfies Eq.~(\ref{A}). This suggests a breakdown of the theory. Nevertheless, we can consider such models by assuming that this breakdown occurs sometime in the future. In contrast, for $\epsilon<0$, this breakdown does not occur. Instead, as the Universe expands, the value of $\barT$ decreases until it reaches a minimum. In either case, this implies that there is an asymptotic upper limit on $\xi$ as seen in Eq.~(\ref{diff X}), given by
\begin{equation}\label{x limit}
    x_0>
    \begin{cases}
        (-\epsilon)^{1/(1-\epsilon)} &\epsilon<0\;,\\
        [(1-2\epsilon)\epsilon]^{1/(1-\epsilon)} &\epsilon>0\;.
    \end{cases}
\end{equation}
The upper bound on $\xi$ can be calculated by combining Eqs.~(\ref{lambda}) and (\ref{xi}) evaluated in the limit of Eq.~(\ref{x limit}). These limiting  $\xi_\text{lim}$ are listed in Tab.~\ref{tab:DataTableCorr}. 

\begin{figure}
    \centering
    \includegraphics[width=1\linewidth]{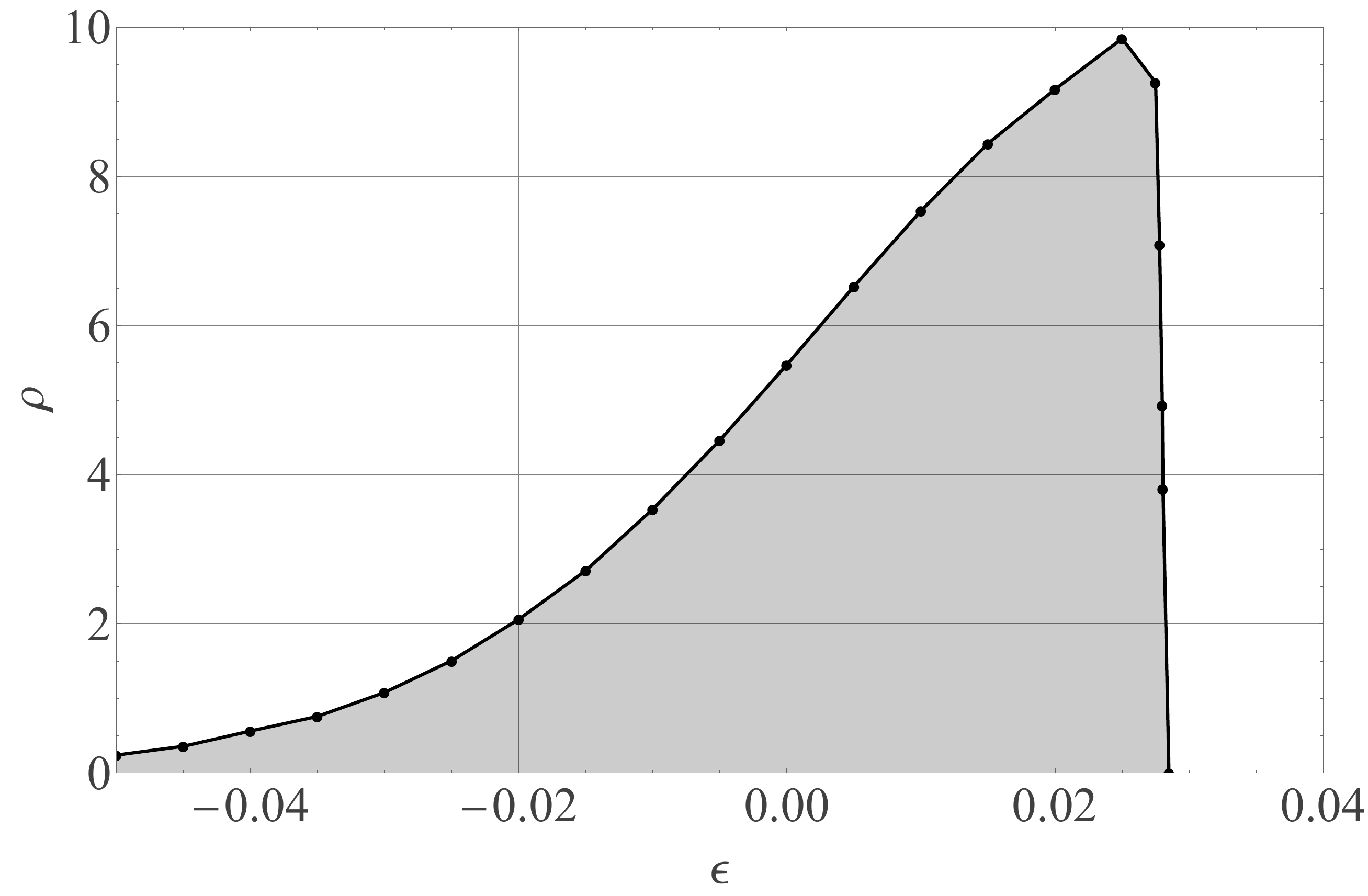}
    \caption{Probability neighborhood for $\epsilon$: The graph has been normalized such that $\int\rho(\epsilon) d\epsilon=1$.}
    \label{fig:Distribution}
\end{figure}

This leads to the probability density
\begin{equation}
    \rho(\epsilon)\propto\frac{\exp\left(-\frac{1}{2}\chi^2_\text{min}\right)}{2\sqrt{\det{\alpha}}}\left[1+\text{erf}\left(\frac{\xi_\text{lim}-\xi_\text{min}}{\sqrt{2}\,\sigma_\xi}\right)\right]\;.
\end{equation}
The resulting probability density is shown in Fig.~\ref{fig:Distribution}. Notice the precipitous decrease of $\rho(\epsilon)$ beginning around $\epsilon=0.027$. This is due to $\xi_\text{min}$ approaching and then exceeding $\xi_\text{lim}$, so that the prior restriction on $\xi$ causes a strong suppression of $\rho(\epsilon)$. The $68\%$ probability range for $\epsilon$ is $\epsilon=0.010^{+0.013}_{-0.021}$.
\vspace{0.2in}
\section{Conclusion}\label{Sec:Conclusion}

Assuming a flat Universe, we have studied the likelihood distribution in $\epsilon$ for $f(R,T)=R+\lambda T^\epsilon$. We expanded our analysis from previous work \cite{Siggia_2025,Siggia_2026} to include updated data from Pantheon+SH0ES and also included measurements from the cosmic microwave background (CMB), baryon accoustic oscillations (BAO) and cosmic chronometer measurements (CC). In \cite{Siggia_2026}, we found that all negative values of $\epsilon$ worked well, even $\epsilon = - \infty$, and they all produced marginally smaller values of $\chi^2$ compared to $\Lambda$CDM. In this paper we find that $\epsilon$ is highly constrained, with a value of $\epsilon=0.010^{+0.013}_{-0.021}$ at 68\% confidence level. The tighter constraints mostly are due to the improved and expanded supernova data from Pantheon+SH0ES and the inclusion  of the DESI DR2 data. 

We note that $\epsilon = 0$, which corresponds to the standard $\Lambda$CDM model, lies well within the tight error bars. This suggests there is at present no preference for this modification of gravity to the form $f(R,T) = R + \lambda T^\epsilon$.

\section*{Data Availability}

The data supporting the findings of this article are openly available and can be found most easily at \cite{Chen_2019,DESI_DR2,GameOver,10.1093/mnras/stad1621,Brout_doi}. The complete data set of Pantheon+SH0ES \cite{Brout_2022,Riess_2022} can be found at URL: {\href{https://pantheonplussh0es.github.io/}{pantheonplussh0es.github.io/}}\;\cite{Brout_doi}. The code used to analyze these data for our models is openly available at {\href{https://github.com/VRSiggia/Analysis_Cosmo_Data}{https://github.com/VRSiggia/Analysis\_Cosmo\_Data}}.

\medskip

\end{document}